\documentstyle[12pt]{article}
\newtheorem{Lemma}{Lemma}[section]
\newtheorem{Proposition}{Proposition}[section]
\newtheorem{Definition}{Definition}[section]
\newtheorem{Theorem}{Theorem}[section]
\newtheorem{Corollary}{Corollary}[section]

\newcommand {\calA}{\mbox{${\cal A}$}}

\newcommand {\za}{\mbox{${\cal Z}({\cal A})$}}

\begin{document}

\title{State Vector Reduction as a Shadow 
of a Noncommutative Dynamics}
\author{Michael Heller\\
Vatican Observatory, V-00120 Vatican City State
\and Wies{\l}aw Sasin and Zdzis{\l}aw Odrzyg\'o\'zd\'z \\
Institute of Mathematics, \\
Warsaw University of Technology \\
Plac Poitechniki 1, 00-661 Warsaw, Poland}

\maketitle

\begin{abstract}{\rm A model, based on a noncommutative
geometry, unifying general relativity and quantum mechanics, is
further developed.  It is shown that the dynamics in this model
can be described in terms of one-parameter groups of random
operators.  It is striking that the noncommutative counterparts
of the concept of state and that of probability measure
coincide.  We also demonstrate that the equation describing
noncommutative dynamics in the quantum mechanical approximation
gives the standard unitary evolution of observables, and in the
``space-time limit'' it leads to the state vector reduction. The
cases of the spin and position operators are discussed in
details.}\end{abstract}

\section{{\rm Introduction}} There are many attempts to create a 
quantum theory of gravity, or at least a generalized version of
the theory of general relativity, based on a noncommutative
geometry (for example,
\cite{Chamsed,ConCham,Con,Hajac,Madore,MadMou,MadSae,Sitarz}).  The 
starting idea of all these works consists in transforming the
space-time manifold into a noncommutative space.  In the series
of works (Refs.  12-15) we have proposed another strategy.  It
turns out that if one replaces space-time $M$ with a groupoid $
G$ ``over'' $M$, one can construct a consistent model unifying
general relativity and quantum mechanics.  The idea is to define
a noncommutative algebra $ {\cal A}$ of complex valued functions
on the groupoid $G$ (with convolution as multiplication) such
that the algebra ${\cal A}$, if narrowed to (a subset of) its
center ${\cal Z}({\cal A})$, is isomorphic with the algebra $
C^{\infty}(M)$ of smooth functions on a space-time $M$ (with the
usual pointwise multiplication).  Then a noncommutative
(derivation based) differential geometry is developed in terms
of the algebra ${\cal A}$, and a noncommutative generalization
of general relativity is constructed.  It turns out that, after
this construction is done, the algebra $ {\cal A}$ can be
completed to the $C^{*}$-algebra (it is called {\em Einstein} $
C^{*}${\em -algebra\/}).  The next step is to quantize the
system in close analogy with the standard $ C^{*}$-algebraic
method.  Details of this approach are summarized in Section 2.

Noncommutative geometry, which in this approach is supposed to
model the pre-Planck epoch, is strongly nonlocal with no local
concepts (such as that of space point or time instant) having
any meanings.  In Sections 3 and 4 it is shown that, in spite of
this, in the noncommutative regime, the true albeit generalized
dynamics is available.  It is only during the ``phase
transition'' from the noncommutative geometry to the usual
(commutative) geometry that space, time and other local
structures emerge.  It can be demonstrated that some nonlocal
phenomena, known from quantum mechanics and cosmology, such as
Einstein-Podolsky-Rosen experiment
\cite{EPR} or the horizon problem \cite{Essay}, can be explained as 
``shadows'' or remnants of the primordial totally global regime.
It also turns out that in the noncommutative regime the
singularity concept loses its meaning (the algebra ${\cal A}$
does not distinguish between singular and nonsingular states),
and classical singularities are but products of the transition
to standard physics of the commutative era
\cite{sing1,sing2}.  

The main goal of the present paper is to show that the
probabilistic character of quantum mechanics is a direct
consequence of our model, and that the generalized
noncommutative dynamics unifies in itself both the unitary
evolution of quantum observables and the ``reduction of the
state vector''.  The key point is that the von Neumann algebra
${\cal M}$, generated by the algebra ${\cal A}$, is both a
``dynamical object'' and a noncommutative generalization of the
probability measure
\cite{Connes,ConRov}.  On the strength of the Tomita-Takesaki theorem 
we introduce one-parameter groups which serve to define the
generalized evolution of von Neumann operators (i.  e.  elements
of ${\cal M}$).  We demonstrate that these operators correspond
to random operators on a Hilbert space and that generalized
dynamics of these operators, in the quantum mechanical
approximation, gives the unitary evolution of observables known
from quantum mechanics (Section 5), and in the ``space-time
limit'' it leads to the reduction of the state vector; the cases
of the spin and position operators are analyzed in details.
(Section 6).  It looks as if the source of the ``measurement
problem'' in quantum mechanics was the fact that quantum
processes do not occur in space-time whereas the act of
measurement is, out of its very nature, a spatio-temporal event.
As a byproduct of this analysis we show that the generalized
dynamics of our model need not be assumed a priori (by
postulating an additional equation describing this evolution, as
it was done in our previous works), but it can be deduced from
the model under rather mild conditions.  In Section 7, we
collect our main results.

\section{A Noncommutative Unification of General Relativity and 
Quantum Mechanics} In this section we briefly summarize our
model in a quasi-axiomatic way; the model presented earlier
(Refs.  12-15) could be considered as a special instant of this
more general scheme.

\vspace{0.3cm} 1.  Let us consider a product $
G=E\times\Gamma$, where $E$ is a smooth manifold (or, more
generally, a differential space of constant dimension
{\cite{HelSas1}), and $\Gamma$ a Lie group acting on $ E$ (to
the right).  $G$ can be given the structure of a smooth groupoid
(see
\cite[p.  99]{Connes}, \cite{Renault}).  We additionally assume that 
there is a subset $\tilde {{\cal Z}}$ of the center $ {\cal
Z}({\cal A})$ of the algebra ${\cal A}$ such that $
\tilde {{\cal Z}}$ is 
isomorphic with $C^{\infty}(M)$ where $M$ is the space-time
manifold (or its differential space generalization).

\vspace{0.3cm}
2. We define the involutive algebra ${\cal A}
=(C_c^{\infty}(G,{\bf C}),+,*,^{*})$ of compactly supported,
complex valued functions on the groupoid $G$ where $` `+$'' is
the usual addition, and the multiplication is defined to be the
convolution
\[(a*b)(\gamma )=\int_{G_q}a(\gamma_1)b(\gamma_
2),\] where $\gamma =(q,qg)\in G,\,\gamma =\gamma_1
\circ\gamma_2$, and $G_q$ is the fiber of $G$ over $
q\in E$; the integral is taken with respect to the (left)
invariant Haar measure. The involution is defined in the
following way: $a^{*}(\gamma )=\overline {a(\gamma^{-1})}$.  Let
us notice that instead $\gamma =(q,qg)$ we can simply write
$\gamma =(q,g)$.

\vspace{0.3cm} 3.  Let ${\rm D}{\rm e}{\rm r}
{\cal A}$ be the set of all derivations of the algebra ${\cal
A}$; it has the structure of a $ {\cal Z}({\cal A})$-module.  We
define the differential algebra $({\cal A},V)$ where $V$ is a $
{\cal Z}({\cal A})$-submodule of ${\rm D}{\rm e} {\rm r}{\cal
A}$ of the form $V=V_E\oplus V_{\Gamma}$ with $V_E$ being the
set of all derivations ``parallel'' to $E$ and $V_{\Gamma}$ the
set of all derivations ``parallel'' to $
\Gamma$.  

\vspace{0.3cm}
4. A metric on V is defined to be a ${\cal Z} ({\cal
A})$-bilinear nondegenerate symmetric mapping $g:V\times
V\rightarrow {\cal A}$. For our model we choose the metric
\begin{equation}g=pr^{*}_Eg_E+pr_{\Gamma}^{*}
g_{\Gamma}\label{metric}\end{equation} where $g_E$ and
$g_{\Gamma}$ are metrics on $ E$ and $\Gamma$, respectively, and
$pr_E$ and $pr_{\Gamma}$ are the obvious projections. It has
been demonstrated by Madore and Mourad \cite{MadMou} that for a
broad class of derivation based differential algebras the metric
is essentially unique. This is the case for the $\Gamma$-part of
metric (\ref{metric}), but the $ E$-part of this metric is
determined by the Lorentz metric on the space-time $ M$.

\vspace{0.3cm} 5.  Now, we develop the noncommutative differential 
geometry as in \cite{HSL,Towards}:  we define the linear
connection (with the help of the Koszul formula), the curvature
and the Ricci operator ${\bf R}:V\rightarrow V$, and we write
the {\em noncommutative Einstein equation \/}
\begin{equation}{\bf G}=0\label{R2}\end{equation}
where ${\bf G}={\bf R}+2\Lambda {\bf I}$ with $ {\bf R}$ being
the Ricci operator, $\Lambda$ a constant related to the usual
cosmological constant, and $ {\bf I}$ the identity operator;
ker{\bf G} is evidently the solution of this equation.  Because
of the form of metric (\ref{metric}) eq.  (\ref{R2}) can be
written in the form
\[{\bf G}_E+{\bf G}_{\Gamma}=0\]
where ${\bf G}_E$ is the part parallel to $E$, and $ {\bf
G}_{\Gamma}$ is the part parallel to $\Gamma$.  Since in the
$\Gamma$-direction there is essentially one metric, the equation
${\bf G}_{\Gamma}=0$ should be solved for derivations $ v\in
{\rm k}{\rm e}{\rm r}{\bf G}_{\Gamma}\subset V_{\Gamma}$.  The
equation ${\bf G}_E=0$ is a ``lifting'' of the usual Einstein
equation in the space-time $M$ (therefore, it should be solved
for the metric); all derivations $ v\in V_E$ satisfy it, and all
derivations $v\in V_{\Gamma}$ satisfy it trivially.  It can be
easily seen that ${\rm k}{\rm e}{\rm r}{\bf G} ={\rm k}{\rm
e}{\rm r}{\bf G}_E\oplus {\rm k} {\rm e}{\rm r}{\bf G}_{\Gamma}$
is a \za-submodule of V (see
\cite{HSL,Towards}).  

\vspace{0.3cm} 6.  Let $\pi_q:\calA\rightarrow 
{\cal B}({\cal H})$ be a representation of the algebra ${\cal
A}$ in the Hilbert space ${\cal H} =L^2(G_q)$, where ${\cal
B}({\cal H})$ denotes the algebra of bounded operators on ${\cal
H}$, given by the formula
\begin{equation}(\pi_q(a)\psi )(\gamma )=(a_q
*\psi )(\gamma ),\label{Conrepr}\end{equation} $a_q$ is here a
restriction of $a\in {\cal A}$ to the fiber $ G_q,\,$$q\in E$,
and $\gamma\in G,\psi\in {\cal H}$.  The completion of \calA\
with respect to the norm
$\parallel a\parallel\,={\rm s}{\rm u}{\rm p}_{
q\in E}\parallel\pi_q(a)\parallel$ is a $C^{*}$-algebra  (see
\cite[p. 102]{Connes}). This algebra will be called {\em
Einstein} $C^{*}${\em -algebra}.
 
\vspace{0.3cm} 7.  We quantize the above system with the help of the 
algebraic method based on classical works by Jordan, von
Neumann, and Wiener \cite{Jordan}, Segal \cite{Segal1,Segal2},
Haag and Kastler
\cite{Haag}.  Let ${\cal S}$ denote the set of states on the Einstein $
C^{*}$-algebra ${\cal A}$.  We assume that elements of ${\cal
S}$ represent states of the system and pure states of ${\cal S}$
represent pure states of the system.  Let $ a\in {\cal Z}({\cal
A})$ be a Hermitian element of ${\cal A}$ and let $\varphi
\in {\cal S}$.  In such a case, $\varphi (a)$ is the 
expectation value of the observable $a$ if the system is in the
state $
\varphi$.  

Let us, for simplicity, assume that $\Gamma$ is a compact group
(general case is discussed in \cite{Towards}).  Two fibres $
G_p$ and $G_q$ of $G$, $p,q\in E$, are said to be {\em
equivalent\/} if there is $ g\in\Gamma$ such that $q=pg$.  The
set of all functions of ${\cal A}$ which are constant on the
equivalence classes of fibres of this equivalence relation are
called {\em projectible functions\/}; they form a subalgebra of
${\cal A}$ denoted by $ {\cal A}_{proj}$.  It can be easily seen
that ${\cal A}_{proj}\subset {\cal Z}({\cal A} )$, and
consequently ${\cal A}_{proj}$ is a commutative algebra.  In
fact, it is isomorphic with the algebra $C^{
\infty}(M)$ of smooth functions 
on the space-time $M$.  In this way, we recover the usual
general relativity (in Geroch's formulation \cite{Geroch}).  In
subsequent sections we shall show that the standard quantum
mechanics is also incorporated into our model.

We have computed the above presented model for the case in which
$G=E\times D_4$ where $E$ is the total space of the frame bundle
over the Minkowski space-time and $D_4$ a group of 4 rotations
and 4 reflections (it is a noncommutative subgroup of SU(2))
\cite{Towards} and, more generally, for the case when $\Gamma$
is a finite group \cite{HSL}.  These cases should be regarded as
``toy models'' demonstrating the consistency of our approach.

\section{Random Operators} Let ${\cal A}=C_c^{
\infty}(G,{\bf C})$ (in fact, in what follows 
we can assume that ${\cal A}$ is the Einstein algebra).  For
each $ a\in {\cal A}$ there is a function $\rho_a$ on $E$ with
values in the space of operators given by
\[\rho_a(p)=\pi_p(a)\]
for $p\in E$. The function $\rho_a$ is said to be $
\Gamma$-invariant if, for every 
$g\in\Gamma ,\;$$\rho_a(pg)=\rho_a(p)$.

\begin{Lemma}
The function $\rho_a$ is $\Gamma$-invariant if and only if $a\in
{\cal A}_{proj}$.
\end{Lemma}
{\bf Proof.}  Let us notice that $\rho_a(pg)=
\rho_a(p)$ is 
equivalent to $a_{pg}*\xi_{pg}=a_p*\xi_p$ which gives $
a_{pg}=a_p.$$\Box$

\begin{Lemma}
If $\rho_a=\rho_b$ then $a=b$ (almost everywhere).  $
\Box$
\end{Lemma}

Therefore, we have two equivalent descriptions of our
noncommutative geometry:  one in terms of the algebra of smooth,
compactly supported, complex valued functions (with convolution
as multiplication) on the groupoid $G$; another in terms of the
algebra of operator valued functions on $E$.  The first
description is, in many cases, easier to work with, and gives us
the direct contact with better known commutative functional
algebras (such as the algebra $C^{\infty}(M)$ on a manifold
$M$); the second description gives us better insight when the
underlying space is strongly singular.

Let us notice that the groupoid $G=E\times\Gamma$ has the
natural structure of foliation; this foliation will be denoted
by $ {\cal F}$.  The fibres $G_p,\,p\in E,$ are leaves of the
foliation ${\cal F}$ (for simplicity, we assume that $ G$ is a
smooth manifold).  Let the space of leaves be denoted by $ Y$.
We should notice that $Y$ is in the bijective correspondence
with $ E$.  Let $\lambda :G\rightarrow Y$ be the natural
projection of the element $
\gamma\in G$ onto the leaf 
containing $\gamma$, i.  e.,  $\lambda (\gamma )=G_p$ where
$\gamma =(p,g)$; we shall also write $p={\rm b}{\rm e}{\rm
g}\gamma$.

Let us consider a ``bundle'' of Hilbert spaces $ (L^2(G_{\lambda
(\gamma )}))_{\gamma\in G}$. Sections of this bundle form
one-parameter families $(\xi_{
\gamma})_{\gamma\in G}$ such that, for every 
$\xi_{\gamma}$, $\xi_{\gamma}\in L^2(\lambda (\gamma ))$.

\begin{Definition}
The {\em random operator\/} is a one-parameter family
$r=(r_p)_{p\in E}$ of operators $r_p\in {\rm E} {\rm n}{\rm
d}(L^2(G_p)),\,p\in E$, satisfying the following measurability
condition:  for any sections $ (\xi_{\gamma})_{\gamma\in G}$ and
($\eta_{\gamma})_{\gamma\in G}$ of $(L^2(G_{\lambda (\gamma
)}))_{\gamma\in G}$ the function $G\rightarrow {\bf C}$, given
by $\gamma\mapsto (\pi_{{\rm b}{\rm e}{\rm g}(\gamma 
)}(a)\xi_{\gamma},\eta_{\gamma}),$ is measurable (in the usual
sense).
\label{def1}
\end{Definition}
This definition is an application of the general concept of
random operator \cite[p. 51]{Connes} to our case.

The norm of the random operator $r$ is defined as $
\parallel r\parallel ={\rm s}{\rm u}{\rm p}_{
p\in E}\parallel r_p\parallel$.  The equivalence classes of
random operators, modulo almost everywhere, equipped with the
obvious algebraic operations, form the von Neumann algebra
\cite[p.  52]{Connes} which will be denoted by $ {\cal N}$.  It
is also called the {\em von Neumann algebra of the foliation\/}$
{\cal F}$.

Let us notice that any function $a\in {\cal A}$ determines the
random operator $\rho_a$ given by $\rho_a=\pi_p(a)$ for $p\in
E$ (but not every random operator must be determined by a
function $a\in {\cal A}$); $\rho_a$ is in fact a one-parameter
family of operators parametrized by the elements of the set $E$,
and it is easy to check that it satisfies the condition of
Definition \ref{def1}.

\begin{Proposition}
$\bigoplus_{p\in E}\pi_p({\cal A})^{}$ is a subalgebra of the
von Neumann algebra ${\cal N}$ of the foliation ${\cal F}$, and
$ (\bigoplus_{p\in E}$$\pi_p({\cal A}))^{\prime
\prime}$ is a von Neumann subalgebra of 
${\cal N}$. $\Box$
\label{prop1}
\end{Proposition}

\section{State Dependent Evolution of Random Operators} 
First, let us remind some well known concepts.  Let $ A$ be a
$*-$subalgebra of the algebra ${\cal B}({\cal H} )$ of bounded
operators on a Hilbert space.  A vector $\xi\in {\cal B}({\cal
H})$ is {\em separating\/} for the algebra $ A$ acting on ${\cal
H}$ if, for every $T\in A$, $T\xi =0$ implies $T= 0$.  A vector
$\xi\in {\cal B}({\cal H})$ is {\em cyclic \/} for $A$ acting on
${\cal H}$ if $A\xi$ is dense in $ {\cal H}$.  The fact that
$\xi$ is cyclic for $ A$ acting on ${\cal H}$ implies that $\xi$
is separating for $ A'$.  If $A$ is a von Neumann algebra the
reverse is also true (see \cite[Appendix A14]{Dixmier}). Now we
go back to our case.

\begin{Lemma}
If ${\cal A}$ is an algebra with unit then the vector $
\xi ={\bf 1}\in L^2(G_q)$ is cyclic for 
the von Neumann algebra $(\pi_q({\cal A}))^{\prime
\prime}$ acting on $L^2(G_q)$.
\end{Lemma}

{\bf Proof.} We have the obvious equality: ${\cal A}_
q=\pi_q({\cal A}){\bf 1}$ where ${\cal A}_q=\{a_q:a\in {\cal
A}\}.$ The functional space $ {\cal A}_q$ contains polynomial
functions, therefore ${\cal A}_q$ is dense in $L^2(G_q)$, and
consequently $
\xi$ is cyclic in $L^2(G_q)$.

Let $T=\pi_q(a),\,a\in {\cal A}$, be an operator such that $
T\xi =0$. In such a case, $a_q*{\bf 1}=0$ which implies
$a_q\cdot {\bf 1} =0$, and $T=0.$ $\Box$

\begin{Corollary}
If ${\cal A}$ is the algebra with unit then the vector $
\xi =({\bf 1}_q)_{q\in E}$, where 
${\bf 1}_q\in L^2(G_q)$ and ${\bf 1}_q$ is the constant function
equal to one, is cyclic and separable for the von Neumann
algebra ${\cal M} =(\bigoplus_{q\in E}\pi_q({\cal
A}))^{\prime\prime}$ acting on $\bigoplus_{q\in E}L^2(G_q)$.
$\Box$ \end{Corollary}

This allows us to formulate the following theorem.

\begin{Theorem}
Let ${\cal A}$ be the algebra with unit. With every state $
\varphi$ on the  
von Neumann algebra ${\cal M}=(\bigoplus_{q\in E}\pi_q({\cal
A}))^{\prime\prime}$ there is associated the one-parameter group
$(\alpha^{\varphi}_t)_{t\in {\bf R}}$ of automorphisms of ${\cal
M}$ given by
\begin{equation}\alpha^{\varphi}_t(b)=\bigtriangleup^{
-it}b\bigtriangleup^{it},\label{modular}\end{equation} for every
$b\in {\cal M}$, where $\bigtriangleup =S^{*}S$, and the
operator $S\!:\,$${\cal M}\rightarrow {\cal M}$ is defined by
$S(b)(\xi_{\varphi})=b^{*}(\xi_{\varphi})$ where $
\xi_{\varphi}=(\xi_{\varphi_q})_{q\in E}$ is cyclic in $
L^2(G_q)_{q\in E}$. $\Box$
\end{Theorem}

{\bf Proof.} On the strength of Theorem 1 from \cite{HS98} there
exists the unique state $\varphi =(\varphi_q)_{q\in E}$, where $
\varphi_q=(\pi_q(a)\xi_q,\xi_q)$ with $\xi_q$ cyclic in 
$L^2(G_q)$, such that
\begin{equation}\pi_{\varphi_q}(a)[b]=[\pi_q(
a)(b)],\label{GNS}\end{equation} $b\in\pi_q{\cal A}$, is a GNS
representation of the algebra $ {\cal A}$.  Here $[...]$
represents the element of the quotient space with respect to the
ideal $N_{\varphi_q}=\{a\in {\cal A}:\varphi_ q(aa^{*})\}$.
Now, the Tomita-Takesaki theorem
\cite{TT} asserts that the mapping $\alpha^{\varphi}_
t:{\cal M}\rightarrow {\cal M},\,t\in {\bf R} ,$ is a
one-parameter group of automorphisms of the von Neumann algebra
$ {\cal M}$ with the desired properties.  $\Box$

The one-parameter group $\alpha^{\varphi}_t,\, t\in {\bf R}$, is
called the {\em modular group\/} of the state $\varphi$ on the
von Neumann algebra ${\cal M}$ which, by Proposition
\ref{prop1}, is obviously a von Neumann subalgebra of the van 
Neumann algebra ${\cal N}$ of the foliation $ {\cal F}$.

With any element $a\in {\cal A}$ there is associated the random
operator $\rho_a=(\pi_q)_{q\in E}$. If $\rho_a$ belongs to the
von Neumann algebra $ {\cal M}$ we can consider the function of
the form
$t\mapsto\alpha^{\varphi}_t(\rho_a)$
which defines a one-parameter group of random operators
representing the ``evolution'' of these operators starting from
the ``initial'' random operator
$\rho_a=\alpha^{\varphi}_0($$\rho_a)$.  Let us take the closer
look at this evolution.

If we assume that $\rho_a\in {\cal M}$, we have
\[\alpha^{\varphi}_t(\rho_a)=e^{-it\bigtriangleup}
\rho_ae^{it\bigtriangleup}.\]
After differentiating and multiplying by $i\hbar$ this equation
assumes the form
\begin{equation}i\hbar\frac d{dt}|_{t=0}\alpha^{
\varphi}_t(\rho_a)=[\rho_a,-\hbar\ln\bigtriangleup 
].\label{eq1}\end{equation} The ``Hamiltonian''
$-\hbar\ln\bigtriangleup$, through the dependence on the
endomorphism $S$, depends on the state $\varphi$. This equation
should be regarded as describing a noncommutative dynamics of
random operators.

For any random operator $r=(r_q)_{q\in E}$ we define its {\em
eigenfunction $\kappa (q),\,q\in E$\/} by the equation
\begin{equation}r_q\xi =\kappa (q)\xi\label{eq6a}\end{equation}
for any $\xi\in L^2(G_q)$ (for simplicity, we consider a
nondegenerate case).  We also define the {\em eigenfunction}
$\kappa :E\times {\bf R}\rightarrow {\bf C}$
for the ``evolution'' of a random operator $r$ by the following
equation
\begin{equation}(\alpha^{\varphi}_{\tau}(r))_{
q\in E}\xi =\kappa (q,t)\xi ,\label{eq1a}\end{equation} for any
$\xi\in L^2(G_q)$ where $\alpha^{\varphi}_{
\tau}(r)$ is a random operator at the instant 
$t=\tau$.

In the noncommutative regime there is no time, and consequently
there cannot be any dynamical equations (in the usual sense).
However, as we have seen, a noncommutative counterpart of
dynamics is encoded in (state dependent) modular groups of
random operators.  It is important to see how these modular
groups project to the standard dynamics in the quantum
mechanical case.

First, let us notice that if a random operator is of the form
$r=$$(\pi_q(a))_{q\in E}$, where $a\in {\cal A}_{ proj}$, then
the function $\kappa :E\rightarrow {\bf C}$ is
$\Gamma$-invariant, i.  e.,  $\kappa (qg)=\kappa (q)$ for every
$ g\in\Gamma$.

\begin{Lemma}
If $a\in {\cal A}_{proj}$ then the operator $
\rho_a$ can be identified with the 
function $\rho_a:M\rightarrow {\bf C}$.  For any $ x\in
M,\;\kappa (x)$ is the eigenvalue of the operator $\rho_a(p)$
where $\pi_M(p)=x$.
\label{lem2}
\end{Lemma}

{\bf Proof.}  We have
\[\rho_a(p)\xi =a_p*\xi .\]
Since $a\in {\cal A}_{proj}$, $a_p=$const on the set $
\pi_M^{-1}(x)$. For $x\in M$, such that 
$\pi_M(p)=x$, one has $a_p=\kappa (x)$ where
$\kappa :M\rightarrow {\bf C}$
is a function on $M$ such that $a=\kappa\circ
\pi_M$. Therefore,
\begin{equation}\rho_a(p)\xi =\kappa (x)\cdot
\xi .\label{eigen1}\end{equation}
Consequently, $\kappa (x)$ is the eigenvalue of the operator $
\rho_a$ and we can 
identify $\rho_a(p)$ with $\kappa$. $\Box$

{\bf Corollary.} {\em For} $a\in {\cal A}_{pr oj}$ {\em the
operator} $\pi_p(a)=\rho_a(p),\, p\in E${\em , is a homothety
with the constant} $\kappa (x)${\em , and consequently its
eigenspace is the whole space} $L^2(G_p)${\em .} $\Box$

The function $\rho_a$ is, in fact, the spectrum of the operator
$ a$.  If $\rho_a$ is a random operator, the function
$\rho_a:M\rightarrow {\bf C}$ is measurable in the usual sense
(because of the measurability condition in Definition
\ref{def1}) The function $\kappa$ (or $\rho_a(p)$ understood as
a function on $ M$) is an eigenfunction of $a$.  Of course, if
$a$ is Hermitian, the eigenvalues $
\kappa (x)$ are real.  

Lemma \ref{lem2} is expressed in terms of operator valued
functions on $E$. However, it can be equivalently expressed in
terms of the algebra ${\cal A}=C^{\infty}(G,{\bf C})$. It then
says that with every $a\in {\cal A}_{proj}$ there is the
canonically associated function (measurable in the usual sense)
$\tilde { a}:{\cal M}\rightarrow {\bf C}$ such that $\tilde {
a}\circ\pi_M=a$.  For any $x\in M$, $a(x)$ is an eigenvalue of
the operator $
\rho_a(q)$ where 
$\pi_M(q)=x$.

Let ${\cal M}=(\bigoplus_{q\in E}\pi_q({\cal A}
))^{\prime\prime}$ be the von Neumann subalgebra of the von
Neumann algebra ${\cal N}$ of the foliation $ {\cal F}$. It is
easy to check that the mapping
$\rho :{\cal A}\rightarrow {\cal M}$ given by
$\rho (a)=\rho_a$,
for $a\in {\cal A}$, is a homomorphism of algebras, and
consequently we have
$\rho ({\cal Z}({\cal A}))\subset {\cal Z}({\cal M}
)\subset {\cal Z}({\cal N})$. It follows that if $a\in {\cal
A}_{proj}\subset {\cal Z}({\cal A})$ then the one-parameter
group $
\alpha^{\varphi}_t(a)$ 
is constant.  Therefore, if we go to the space-time
approximation (if we restrict to ${\cal A}_{proj}$) the
noncommutative dynamics is switched off.  Let us notice,
however, that this is valid only for a given state $
\varphi$.  
It is not unlike in the Schr\"odinger picture of quantum
mechanics in which operators are constant and all time
dependence goes to the state vectors. We should expect that the
dynamics reappears in the quantum mechanical approximation.

Such an approximation is obtained if we narrow the algebra $
{\cal A}$ to its subalgebra
\[{\cal A}_{\Gamma}:=\{f\circ pr_{\Gamma}:f\in
fC^{\infty}_c(\Gamma ,{\bf C})\}\] where
$pr_{\Gamma}:G\rightarrow\Gamma$ is the obvious projection.  For
any $ a\in {\cal A}_{\Gamma}$, the random operator
$\rho_a=(\pi_q(a))_{q\in E}$ is a family of operators which can
be identified with each other (because of the natural
isomorphism of leaves of the foliation ${\cal F}$).  In this
sense any random operator $
\rho_a$, with $a\in {\cal A}_{\Gamma}$, is a 
constant family projectible to the ``typical leaf'' $
\Gamma$.  In such a case, 
the operator, to which the random operator $\rho_ a$ projects,
will be denoted by $a_{\Gamma}$; it belongs to ${\rm E} {\rm
n}{\rm d}(L^2(\Gamma ))$.  Let us notice that $ a_{\Gamma}$ is
not a random operator since random operators are defined on the
foliated space and not on the ``typical leaf''.  Now, eq.
(\ref{eq1}) assumes the form
\begin{equation}i\hbar\frac d{dt}|_{t=0}\alpha^{
\varphi}_t(a_{\Gamma})=[a_{\Gamma},-\hbar\ln\bigtriangleup 
f].\label{eq2}\end{equation} The modular group
$\alpha^{\varphi}_t(a_{\Gamma} )$ is here defined with respect
to the von Neumann algebra $(\pi_q({\cal
A}_{\Gamma}))^{\prime\prime}$ where $ q$ is any element of $E$.
Eq.  (\ref{eq2}) describes the evolution depending on the state
$
\varphi$; we shall return to 
this problem in the subsequent section.

It is interesting to notice that the modular group $
\alpha^{\varphi}_t,\,t\in {\bf R}$, determines 
the derivation $v\in {\rm D}{\rm e}{\rm r}{\cal M}$ of the von
Neumann algebra $ {\cal M}$. We define
\[v(\pi_{\varphi_q}(a)):=\frac d{dt}(\alpha^{
\varphi}_t(\pi_{\varphi_q}(a)))\]
where $a\in {\cal M}$, and the representation $
\pi_{\varphi_q}$ is defined by eq. (\ref{GNS}). 
After simple calculations (see \cite{HS98}), from eq.
(\ref{modular}$ )$ we obtain
\[v(\pi_{\varphi_q}(a))=i[\pi_{\varphi_q}(a),\ln
\bigtriangleup ]=i{\rm a}{\rm d}_{\ln\bigtriangleup}
(\pi_{\varphi_q}(a)).\] The one-parameter groups
$\alpha^{\varphi}_t,\, t\in {\bf R}$, for which there exists a
derivation $v\in {\rm k}{\rm e}{\rm r}{\bf G}$ such that
\[v(\pi_{\varphi_q}(a))=\frac d{dt}(\alpha^{\varphi}_
t(\pi_{\varphi_q}(a))\] deserve to be called {\em integral
curves\/} of the noncommutative Einstein equation.

\section{Unitary Evolution of Random Operators}\label{StateInd}
In this Section we show that eq. (\ref{eq1}), in the quantum
mechanical approximation, gives the usual unitary evolution of
quantum observables.

Let us first notice that the (above defined) homomorphism of
algebras $\rho\!:\,{\cal A}\rightarrow {\cal M}$ is a
monomorphism.  Indeed, if $
\rho (a)=0$ then $(\pi_q(a))_{q\in E}=0$ 
which implies that $a_q=0$ for each $q\in E$, and this means
that $ a=0$.  Therefore, we have proved the following lemma.

\begin{Lemma}
$\rho\!:{\cal A}\rightarrow\rho ({\cal A})$ is an isomorphism of
algebras. $
\Box$
\label{lem4.2}
\end{Lemma}

Let ${\cal U}=\{u\in {\cal A}:uu^{*}=u^{*}u=1
\}$ be the unitary group of the algebra ${\cal A}$. 
Then $\rho ({\cal U})$ is the unitary group of the algebra $
\rho ({\cal A})$. Let us notice that 
for the subalgebra ${\cal A}_{\Gamma}\subset {\cal A}$ we have
$\rho ({\cal A}_{\Gamma})\subset
\rho ({\cal A})$ and the unitary group 
of this subalgebra is of the form
\[{\cal U}_{\Gamma}=\{u\in {\cal A}_{\Gamma}:
uu^{*}=u^{*}u=1\}.\] Evidently ${\cal U}_{\Gamma}\subset {\cal
U}$, and correspondingly $
\rho ({\cal U}_{\Gamma})\subset\rho ({\cal U}
)$.

Let ${\cal R}=(\rho ({\cal A}_{\Gamma}))^{\prime
\prime}$ be the von Neumann algebra generated by the algebra 
of random operators $\rho ({\cal A}_{\Gamma})$.  In agreement
with the general construction \cite{ConRov,HS98}, the
automorphisms $
\alpha':{\cal R}\rightarrow {\cal R}$ and $\alpha^{
\prime\prime}:{\cal R}\rightarrow {\cal R}$ 
are said to be {\em inner equivalent\/} if there is an element $
u\in\rho ({\cal U}_{\Gamma})$ such that
\[u\alpha^{\prime\prime}(b)=\alpha'(b)u\]
for every $b\in {\cal R}$. The set of equivalence classes of
this relation is called the {\em group of outer automorphisms\/}
and is denoted by $ {\rm O}{\rm u}{\rm t}$${\cal R}$. As well
known, the one-parameter group $\alpha^{
\varphi}_t,\,t\in {\bf R}$, canonically projects 
onto the (nontrivial) one-parameter group $\tilde{
\alpha}_t,\,t\in {\bf R}$, in ${\rm O}{\rm u}
{\rm t}{\cal R}$ which is independent of the state $\varphi$.

From eq.  (\ref{eq1}) it follows that $\tilde{
\alpha}_t$ satisfies the following 
equation in ${\rm O}{\rm u}\mbox{${\rm t}{\cal R}$}$
\[\frac d{dt}|_{t=0}[\alpha_t(a)]=i[[a],[\ln\bigtriangleup 
]]\] where [...] denotes the equivalence class of inner
equivalence. This equation can also be written in the form
\[i\hbar\frac d{dt}|_{t=0}\tilde{\alpha}_t(\tilde {
a})=[\tilde {a},H]\] where $H=-\hbar [\ln\bigtriangleup ]$.
This equation, after being projected to the ``typical leaf''
$\Gamma$, assumes the form
\begin{equation}i\hbar\frac d{dt}|_{t=0}\tilde{
\alpha}_t(\tilde {a}_{\Gamma})=[\tilde {a}_{\Gamma}
,H]\label{eq3}\end{equation} which is the same as eq.
(\ref{eq2}) but now independent of the state $\varphi$.  This
is, in fact, the Schr\"odinger equation in the Heisenberg
picture of quantum mechanics in which operators evolve but state
vectors are time independent.  In this way, by projecting to the
``typical leaf'' $\Gamma$, we recover from our model the unitary
evolution of ordinary quantum operators.

\section{Reduction of the State Vector}

The product structure of the groupoid $G=E\times
\Gamma$ plays the essential 
role in our model.  The ``$E$-component'' of the model is, in
principle, responsible for its gravitational effects, whereas
the ``$\Gamma$-component'' is responsible for quantum mechanical
effects.  In the quantum mechanical approximation we simply
forget about the $E$-component effects.  In this Section we show
that precisely this fact leads to the effect which, from the
$\Gamma$-perspective, looks like the reduction of the state
vector.  First, however, we must do some preparatory work.

Let us consider a function $f:G\rightarrow {\bf C}$ on the
groupoid $ G=E\times\Gamma$.  With fixed $g\in \Gamma $ we
obtain the function $f_ g:E\rightarrow {\bf C}$ given by
\[f_g(p)=f(p,g)\]
for $p\in E$.  We recognize in it the eigenfunction $
\kappa (q)$ of equation (\ref{eq6a}).  This function determines
the one-parameter family of functions $(f_g)_{g\in\Gamma}$.  For
a fixed $ g\in\Gamma$ we obtain the sequence of the values
$(f_g(p))_{p\in E}$ of the function $f$ on the fibre $ E\times
\{p\}$ for each $p\in E.$ In particular, for any Hermitian
element $a\in {\cal A}_{proj}$ we obtain the sequence of real
values $a_g(p),p\in E$. In this case, the sequence
$(a_g(p))_{p\in E}$ does not depend of $g\in\Gamma$ (since
elements of ${\cal A}_{proj}$ are constant on fibres $G_p$ for
every $p \in E$).  On the strength of Lemma
\ref{lem2} and the subsequent corollary, if $
a\in {\cal A}_{proj}$ is Hermitian then the random operator
$\rho_a=(\pi_p(a))_{ p\in E}$ is a one parameter family of
homotheties with constants $a(p,g)$ where $g$ is any fixed
element of the group $\Gamma$, and the operator $r_p=\pi_ p(a)$,
for a fixed $p\in E$, satisfies the eigenvalue equation
\begin{equation}r_p\xi =\kappa (p)\xi\label{eq12}\end{equation}
for $\xi =L^2(G_q).$ The eigenspace of the operator $ r_p$ is
the whole Hilbert space.  Since $a\in {\cal A}_{proj}$ the
eigenfunction $
\kappa (p)=a(p,g)$ assumes 
constant values on the fibres $\pi_{^{}}^{-1} (x),\,x\in M$.
Consequently, there is the real valued function $\tilde{\kappa
}:M\rightarrow {\bf R}$ such that $\tilde{\kappa}\circ\pi_M=
\kappa$, and the random 
operator $\rho_a=(r_p)_{p\in E}$ has the following set of
eigenvalues
\[\{\kappa (p):p\in E\}=\{\tilde{\kappa }(x):
x\in M\}.\]

Let us notice that in every act of measurement the measuring
apparatus is always located at a given point in space-time $
x\in M$.  This automatically causes the function $\tilde{
\kappa}$ to ``collapse'' to its value 
$\tilde{\kappa }(x)$ (in the examples below we shall see that
this indeed is connected with the reduction of the state
vector). Such a procedure is meaningful only with respect to
operators which commute with the position operator since
measuring the eigenvalue $
\tilde{\kappa }(x)$, for a given $x$, 
presupposes the knowledge of $x\in M$.

The above analysis is carried out from the perspective of the
$E$-component of our model; to go back to the standard
measurement interpretation in quantum mechanics we must see how
the process looks like from the perspective of its
$\Gamma$-component.  Let us choose an orthonormal basis
$\{\psi_n(g)\}$ in the Hilbert space $ L^2(\Gamma )$.  We are
looking for the operator $\rho_{\Gamma}$ acting in this space
the eigenvalues of which would be $\kappa (p)$.  This does not
necessarily mean that the spectrum of this operator is
continuous.  For instance, if the function $
\kappa (p)$ is 
constant there is only one eigenvalue.  Let us first assume that
the spectrum $\kappa (p)$ is discrete.  In this case, the looked
for operator is
\[\rho_{\Gamma}=\sum_n^{}\kappa_nP_{\psi_n},\]
where $P_{\psi_n}$ is the projector onto the direction
determined by $
\psi_n$ (for simplicity, we consider the nondegenerate case).
In the case of continuous spectrum, we proceed analogously and
use the corresponding spectral theorem (see below Example 2).
In this way, we recover the standard formalism of quantum
mechanics.  If this formalism is taken separately, without
paying attention to what happens in the $E$-direction, all
interpretative problems of quantum mechanics immediately arise.
The following examples show that these problems are naturally
solved if the model is regarded in its totality.

{\bf Example 1. Spin measurement.} In \cite{EPR} it has been
shown that to the usual z-component spin operator $\hat {S}_z$
there correspond two elements $ s_1$ and $s_2$ of the
noncommutative algebra ${\cal A}$ such that
\[\pi_p(s_1)\psi =+\frac {\hbar}2\psi\;\;{\rm i}
{\rm f}\;\;\psi\in {\bf C}^{+}\] and
\[\pi_p(s_2)\psi =-\frac {\hbar}2\psi\;\;{\rm i}
{\rm f}\;\;\psi\in {\bf C}^{-}\] where ${\bf C}^{+}={\bf
C}+\{0\}$ and ${\bf C}^{ -}=\{0\}+{\bf C}${\bf .}  Since $s_1$
and $s_ 2$ are observables we assume that they are Hermitian and
elements of $ {\cal A}_{proj}$; consequently, they can be
regarded as real valued functions on $ M$ defined by:
$s_1=+\frac {\hbar}2$ and $s_2=-\frac {\hbar} 2$.  Since $s_1$
and $s_2$ are constant functions they also belong to the
subalgebra ${\cal A}_{\Gamma}$.  Both $ s_1$ and $s_2$ are
homotheties, and consequently the entire Hilbert space
$L^2(G_p)$ is the eigenspace of these operators.  This means
that the results of the spin measurements are strictly
predetermined (i. e., they are obtained with certainty),
although at the present we do not know the mechanism of this
predetermination. However, for the sake of concreteness, let us
naively assume that it is given by the following random operator
\begin{equation}r_p=\left\{\begin{array}{ll}
\pi_p(s_1)&\mbox{if $\pi^3(p)\geq 0$,}\\
\pi_p(s_2)&\mbox{if $\pi^3(p)<0,$}\end{array}
\right.\label{random}\end{equation}
where $\pi^3(p)=x^3$ is the projection onto the third space
coordinate ($z$-coordinate).  It is indeed the random operator
since the mappings $\gamma\mapsto (+\frac {\hbar}2\xi_{\gamma},
\eta_{\gamma})$ and $\gamma\mapsto (-\frac {\hbar}2\xi_{\gamma},
\eta_{\gamma})$ are measurable.

To see what happens in the perspective of the observer
performing the measurement we must situate the observer in
space-time $M$ (the $E$-component of our model). Let us suppose
that the measuring apparatus is at a space-time point
$x=\pi_M(p),\,p\in E$.  The result of the measurement will be
$+\frac {\hbar}2$ or $-\frac {\hbar}2$ (with probability 1)
depending on whether $\pi^3(p)$$\geq 0$ or $\pi^3(p)<0$ with
$\pi_M(p)=x$. These two conditions are not known to the
observer, therefore, in computing the probability of the result
he uses the $\Gamma$-perspective (i.  e.,  the standard
machinery of quantum mechanics), and the outcome of the
measurement looks for him as the ``collapse of the wave
function''. We can see this by choosing two orthonormal vectors
$\psi_1 ,\,\psi_2\in L^2(\Gamma )$ which span the subspace
$\langle\psi_1,\psi_2\rangle_{{\bf C}\,}
\subset L^2(\Gamma )$; then with the help of the spectral
theorem we recover the usual spin operator
\[\hat {S}_z=+\frac {\hbar}2P_{\psi_1}-\frac {
\hbar}2P_{\psi_2}\]
where $P_{\psi_1}$ and $P_{\psi_2}$ are projecting operators
onto the directions determined by $\psi_1$ and $\psi_2$,
respecrively.  In this way, the operator $\hat {S}_z$ is
determined by the random operator $r_p$.  As usually, if the
system is in the state $\varphi$ the probability that the result
of a measurement will give $+\frac {\hbar}2$ is $|\langle\varphi
,\psi_1\rangle |^2$, and analogously for $-\frac {\hbar}2$. Of
course, in the act of measurement the state vector $\varphi $
collapses either to the eigenvalue $+\frac{\hbar }{2}$ or to the
eigenvalue $-\frac{\hbar }{2}$ in agreement with the standard
procedure of quantum mechanics.

The conditions ``either $\pi^3(p)\geq 0$ or $\pi^3(p)<0$'' in
the formula (\ref{random}) were put there by hand and we could
easily imagine some other conditions which would do the job.
However, it could be hoped that if the theory is more developed
(i. e., if the concrete algebra ${\cal A}$ and the concrete
group $\Gamma $ are chosen basing on physical grounds), the
correct mechanism selecting either $\pi_p(s_1)$ or $\pi_p(s_2)$
will be determined by the theory itself. By now, we could only
guess that this mechanism is connected with the random character
of the operator $r_p$ (formula (\ref{random})). The essential
point is that since $\pi_ p(s_1)$ is a homothety the eigenspace
of the eigenvalue $+\frac {\hbar}2$ is the entire Hilbert
space{\bf ,} and the result of the measurement must be strictly
predetermined.  The same is true for the eigenvalue $-\frac
{\hbar}2$.  This means that there is some ``very well hidden
mechanism'' predetermining the outcome of the measurement.

{\bf Example 2.  Position measurement.}  Let, for simplicity, $
M$ be ${\bf R}^4$, and let $\widetilde{pr}_k:{\bf
R}^4\rightarrow {\bf R}$, $ k=0,1,2,3,$ be the projection
function defined by $ $$\widetilde{pr}$$_k(x^0,x^1,x^2
,x^3)=$$x^k$.  One can see that
$pr_k=\widetilde{pr}\circ\pi_M\circ\pi_E$, ($
\pi_E:G\rightarrow E$ being the obvious projection) 
is a Hermitian element of ${\cal A}_{proj}$.  It can be easily
guessed that $ pr_{\kappa}$ is an observable corresponding to
the position operator.  Its eigenvalue equation is
\begin{equation}\pi_p(pr_k)\xi
=pr_k(x)\xi\label{position}\end{equation} for $\xi\in L^2(G_q)$,
where $x=\pi_M(p)$, $p
\in E$. Hence we obtain
\begin{equation}\pi_p(pr_k)\xi =x^k\xi
,\label{position2}\end{equation} as it should be (the position
operator acts by multiplication).  The spectrum of the position
operator $\pi_p(pr_k )$ is evidently {\bf R}.  The operator
$\pi_p$($pr_k)$ is a homothety, and consequently the entire
Hilbert space $L^2(G_p)$ is the eigenspace corresponding to the
eigenvalue $ pr_k(x)$.  In other words, the result of the
position measurement of a quantum object is always
predetermined, although at the present stage of the development
of the model the mechanism of this predetermination is not known.
Therefore, the only thing we could do is to change to the
$\Gamma$-perspective.  We simply look for the operator acting on
the Hilbert space $L^2(\Gamma )$ the spectrum of which is equal
to {\bf R}.  By using the spectral theorem we find (in the
one-dimensional case)
\[\hat {X}=\int_{-\infty}^{+\infty}xdE(x)\]
where $E$ is a suitable spectral measure (see, for instance,
\cite[pp.24-31]{Omnes}).  Then we can write down the 
standard eigenvalue equation (which essentially is the same as
eq.  (\ref{position2})) and compute the probabilities of the
expected results.  After completing the measurement we would say
that the ``wave function has collapsed''. However, if the entire
model is taken into account there is no real collapse; the
measurement result is strictly predetermined by the fact that
$\pi_p(pr_k)$ is a homothety.

\section{Conclusions} The overview picture that emerges from the 
above analysis is the following.  In general, the noncommutative
regime is atemporal.  The only meaning which we can ascribe to
the term ``dynamics'' is through the fact that certain geometric
quantities are expressed in terms of derivations of the algebra
$ {\cal A}$ defining the considered noncommutative geometry.
Derivations are counterparts of vector fields, and as such they
can be regarded as modelling certain type of ``global change''.
The concept of dynamics improves if the algebra ${\cal A}$ has
properties admitting the existence of one-parameter modular
groups (see \cite{HS98}).  Then the von Neumann algebra,
generated by the algebra ${\cal A}$, becomes a dynamical object
(see
\cite{ConRov}), and modular groups describe the ``evolution'' of the 
corresponding operators.  In what follows, we shall assume that
the algebra ${\cal A}$ admits the existence of modular groups.

In our model this means that the dynamics of the system is
described by eq.  (\ref{eq1}).  This dynamics (or ``evolution'')
depends on the state $\varphi$ in which the system finds itself.
It is especially interesting to consider the evolution of random
operators.  This is not a limitation since, by Lemma
\ref{lem4.2} there is an isomorphism between the algebra ${\cal
A}$ and the von Neumann algebra of random operators.  This fact
has its further important consequences.  Random operators, as
elements of the von Neumann algebra, are probabilistic objects
albeit in a generalized sense.  Let us remind that the
noncommutative counterpart of the probability space is a pair $
({\cal M},\varphi )$ where ${\cal M}$ is a von Neumann algebra
and $
\varphi$ a (faithful and normal) state 
on ${\cal M}$.  By the definition of state, $
\varphi$ is positive (i.  e.  $\varphi (aa^{*}
)\geq 0$ for every $a\in {\cal M}$), and normalized (i.  e.,  $
\varphi ({\bf 1})=1$), in close analogy to the 
standard probability measure.  It is striking that in the
noncommutative regime the concept of state and that of
probability measure coincide.  If we go to the quantum mechanics
approximation these concepts split but remain strictly
interconnected.  We are entitled to say that the probabilistic
character of quantum mechanics is the consequence of the fact
that the quantum mechanical observables are but ``shadows''
(projections) of random operators.

We have demonstrated that in the dynamical equation (\ref{eq1})
there are encoded both the unitary evolution of observables of
the standard quantum mechanics and the reduction of the state
vector (``collapse of the wave function'') occurring in the act
of measurement of a given observable.  To obtain the unitary
evolution of observables of the usual quantum mechanics two
steps must be made.  The first step is to project dynamical
equation (\ref{eq1}) to the ``typical leaf'' $
\Gamma$.  This leads to eq.  (\ref{eq2}).  As the consequence of
this procedure the random operator $\rho_a$ changes into the
ordinary operator $ a_{\Gamma}$, but its evolution is still
state dependent.  In the second step, we form the group of outer
automorphisms (see Section \ref{StateInd}) and obtain
one-parameter groups $\tilde{\alpha}_t,\,t\in {\bf R}$, which
are now state independent.  With this modification dynamical
equation (\ref{eq2}) changes into eq.  (\ref{eq3}) which is the
usual Schr\"odinger equation (in the Heisenberg picture) of
quantum mechanics describing the unitary evolution of
observables.  This process of the transition to quantum
mechanics ``truncates'' the model leaving aside its
$E$-component (in this sense the usual quantum mechanics is
incomplete).  We should notice that precisely the $E$-component
of the model is responsible for all measurements (every
measuring device as a macroscopic object is situated in
space-time).  We have shown that exactly this ``truncation'' is
the reason why what is seen from the $\Gamma$-perspective looks
like a sudden collapse of information in the act of measurement.

We could briefly summarize the situation by saying that the
unitary evolution of quantum observables and the reduction of
the state vector in the act of their measurements are but two
different ``projections'' of the same process, namely, of the
generalized dynamics in the noncommutative regime.

\end{document}